\documentclass[preprint2]{aastex}

\usepackage{graphicx}
\usepackage{url}
\usepackage{color}

\shorttitle{White-light flare height}
\shortauthors{Mart{\' i}nez Oliveros et al.}

\begin{document}

\title{The height of a white-light flare and its hard X-ray sources}

\author{Juan-Carlos Mart{\' i}nez Oliveros$^{1}$, Hugh~S. Hudson$^{1,2}$, Gordon~J. Hurford$^{1,3}$, \\
S{\" a}m Krucker$^{1,3}$, R.~P. Lin$^{1,7}$, Charles Lindsey$^{4}$,\\ Sebastien Couvidat$^5$, Jesper Schou$^5$, and W.~T. Thompson$^6$} 
\affil{$^1$Space Sciences Laboratory, UC Berkeley, CA, USA 94720; \\
$^2$School of Physics and Astronomy, University of Glasgow; \\
$^3$Institute of 4D Technologies, School of Engineering, University of Applied Sciences\\ North Western Switzerland, 5210 Windisch, Switzerland;\\
$^4$North West Research Associates, CORA Division, Boulder, CO USA\\
$^5$W.~W. Hansen Experimental Physics Laboratory, Stanford University, Stanford, CA USA\\
$^6$Adnet Systems, Inc., NASA Goddard Space Flight Center, code 671, Greenbelt MD USA\\
$^7$Physics Department, University of California, Berkeley, CA,\\ 
and School of Space Research, Kyung Hee University, Yongin, Korea.}

\begin{abstract}
We describe observations of a white-light flare (SOL2011-02-24T07:35:00, M3.5) close to the limb of the Sun, from which we obtain estimates of the heights of the optical continuum sources and those of the associated hard X-ray sources.
For this purpose we use hard X-ray images from the \textit{Reuven Ramaty High Energy Spectroscopic Imager (RHESSI)}, and optical images at 6173~\AA~from the \textit{Solar Dynamics Observatory (SDO)}.
We find that the centroids of the impulsive-phase emissions in white light and hard X-rays ($30-80$~keV) match closely in  central distance (angular displacement from Sun center), within uncertainties of order 0.2$''$.
This directly implies a common source height for these radiations, strengthening the connection between visible flare continuum formation and the accelerated electrons.
We also estimate the absolute heights of these emissions, as vertical distances from Sun center.
Such a direct estimation has not been done previously, to our knowledge.
Using a simultaneous 195~\AA~image from the \textit{Solar-Terrestrial RElations Observatory (STEREO-B)} spacecraft to identify the heliographic coordinates of the flare footpoints, we determine mean heights above the photosphere (as normally defined; $\tau  = 1$ at 5000~\AA) of $305 \pm 170$~km and $195 \pm 70$~km, respectively, for the centroids of the hard X-ray (HXR) and white light (WL) footpoint sources of the flare.
These heights are unexpectedly low in the atmosphere, and are consistent with the expected locations of 
$\tau = 1$ for the 6173~\AA~and the $\sim$40~keV photons observed, respectively.
\end{abstract}


\section{Introduction}\label{sec:intro}
The white-light continuum of a solar flare (WLF) was the first manifestation of a flare ever detected \citep{1859MNRAs..20...13C,1859MNRAs..20...16H}.
Nevertheless its origin has remained enigmatic over the intervening centuries.
This continuum contains a large fraction of the total luminous energy of a flare \citep[e.g.,][]{1989SoPh..121..261N}, and so its identification has always posed an important problem for solar and stellar physics.
Although the most obvious flare effects appear in the chromosphere and corona, the physics of the lower solar atmosphere  has great significance for the reasons described by Neidig.

The association of white-light continuum with the impulsive phase of a solar flare has long been known, and cited as an indication that high-energy nonthermal particles have penetrated deep into the lower solar atmosphere \citep{1970SoPh...15..176N,1970SoPh...13..471S}.
Such an association would link the acceleration of ``solar cosmic rays,'' as they were then known, with the intense energy release of the impulsive phase, and this association has proven to be crucial to our understanding of flare physics \citep{1976SoPh...50..153L,2003ApJ...595L..69L,2005JGRA..11011103E,2007ApJ...656.1187F}.
Because the continuum appears in the emission spectrum, and because this nominally originates in an optically-thick source, this energy release may drastically distort the structure of the lower solar atmosphere during a flare.
Accordingly the usual methods for modeling the atmosphere \citep[e.g.,][]{1981ApJS...45..635V}, which assume hydrostatic equilibrium, may be too simple.

\begin{figure*}[htpb]
 \includegraphics[width=0.95\textwidth]{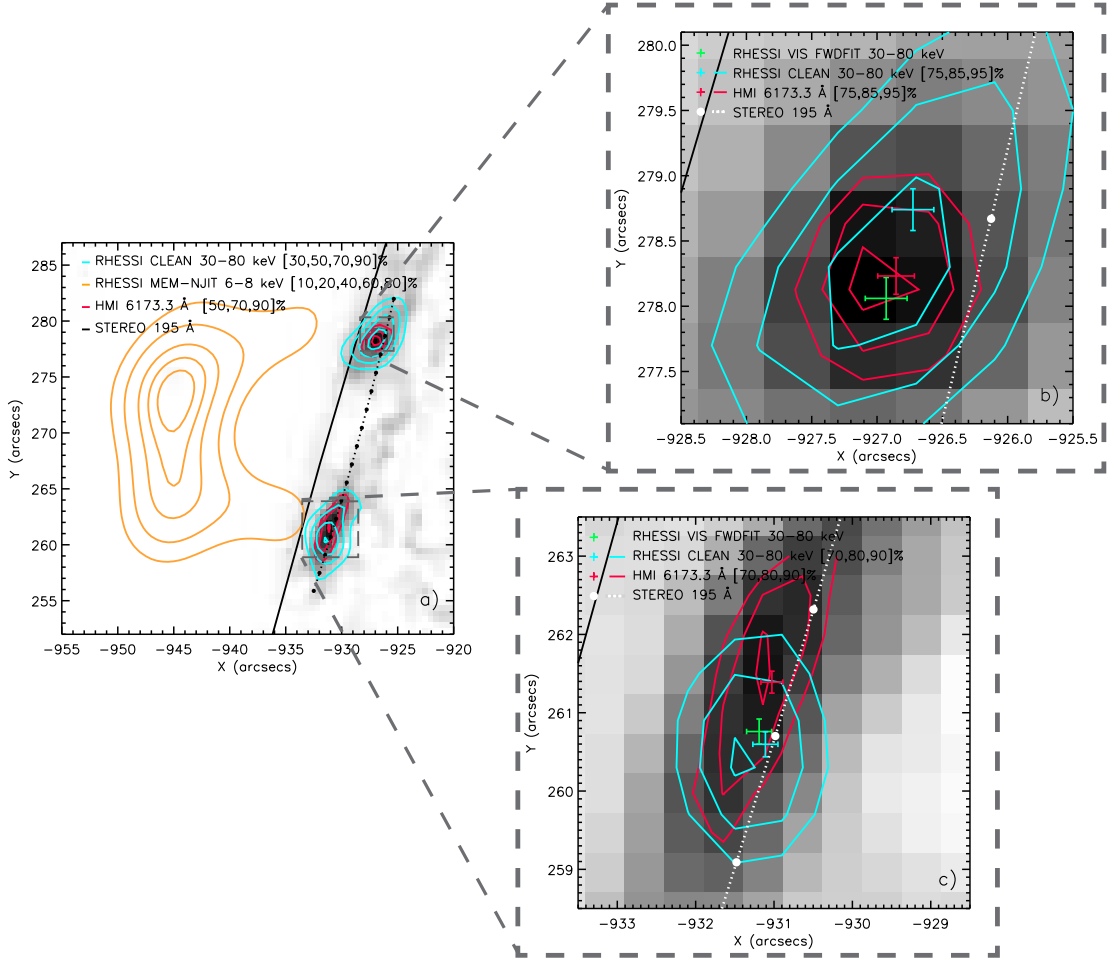}
 \caption{
 HMI intensity continuum difference image (prime, 07:31:13.40~UT; reference, images from 07:25-07:28~UT) combined with white-light difference and RHESSI CLEAN contour plots (red and blue, respectively). The Hard X-ray images (30-80 keV) made with the CLEAN technique for the interval 07:30:50.9 -- 07:31:35.9~UT, exactly that of the HMI cadence. These images were made with \textit{RHESSI} subcollimators 1-4, with uniform weighting, giving an angular resolution (FWHM) of 3.1$''$. The orange contours in panel a) show the 6-8 keV Soft X-ray source at the same time, defining a loop structure connecting the footpoints. The dotted line shows the locus of the STEREO/EUVI source positions, which in this projection show the projected angular location of the photosphere (see Section~\ref{sec:absolute}).}
\label{fig:footpoints}
\end{figure*}

The association with hard X-ray bremsstrahlung has always suggested $>$10~keV electrons in particular \citep{1972SoPh...24..414H,1975SoPh...40..141R,1992PASJ...44L..77H,1993SoPh..143..201N}.
According to the standard thick-target model the primary particle acceleration occurs in the corona above the flare sources.
The presence of fast electrons automatically implicates the chromosphere as well, rather than the photosphere, because of the relatively short collisional ranges of such particles.
Nevertheless the indirect nature of the bremsstrahlung emission mechanism \citep[e.g.,][]{1971SoPh...18..489B} has made it difficult to rule out other processes, such as energy transport by protons \citep{1969sfsr.conf..356E,1970SoPh...15..176N,1970SoPh...13..471S}.
The bremsstrahlung signature may also result from acceleration directly in the lower atmosphere, rather than in the corona \citep{2008ApJ...675.1645F}.
Finally radiative backwarming could also provide a mechanism for explaining for the observed tight correlation of hard X-rays and white-light continuum \citep{1989SoPh..124..303M}.
This mechanism involves irradiation and heating of the photosphere, rather than the chromosphere, as would correspond to the shorter stopping distance of energetic electrons in a thick-target model \citep{1972SoPh...24..414H}.

In the modern era we are seeing a rapid increase in our understanding of these processes, thanks to the excellent new data from various spacecraft and ground-based observatories.
This paper reports on a first good example of a flare near the limb, observed at high resolution both in hard X-rays by \textit{RHESSI}, and at 6173~\AA~in the visible continuum by the \textit{Solar Dynamics Observatory (SDO)} spacecraft via its Helioseismic Magnetic Imager (HMI) instrument \citep{2012SoPh..275..229S}.
This event, SOL2011-02-24T07:35 (M3.5), occurred just inside the limb (NOAA coordinates N14E87), so that source heights could be compared by simple projection.
We analyze data from \textit{RHESSI}, HMI, and the Extreme Ultraviolet Imager (EUVI) on \textit{STEREO-B)}.
\cite{2011A&A...533L...2B} have already studied this flare, using the same data but without reference to the \textit{STEREO-B} observations.

\section{The observations}\label{sec:data}

The flare we study here (Figure~\ref{fig:footpoints}) occurred close to the geocentric east limb.
Both hard X-rays and white-light emission come clearly from the visible hemisphere, as  we establish below via the use of 
STEREO-B/EUVI images. Figure~~\ref{fig:footpoints} shows the HMI intensity continuum difference image at 07:31:13.40~UT,  with the reference image constructed as the average of HMI data from 07:25-07:28~UT.
The white-light contrast and area for this M3.5 flare are consistent with the trends found in surveys \citep{2003A&A...409.1107M,2006SoPh..234...79H,2009RAA.....9..127W} based on \textit{Yohkoh}, \textit{TRACE}, and \textit{Hinode} observations, respectively.

\begin{figure}[htpb]
\includegraphics[width=0.95\columnwidth, trim = 0 0 0 -20]{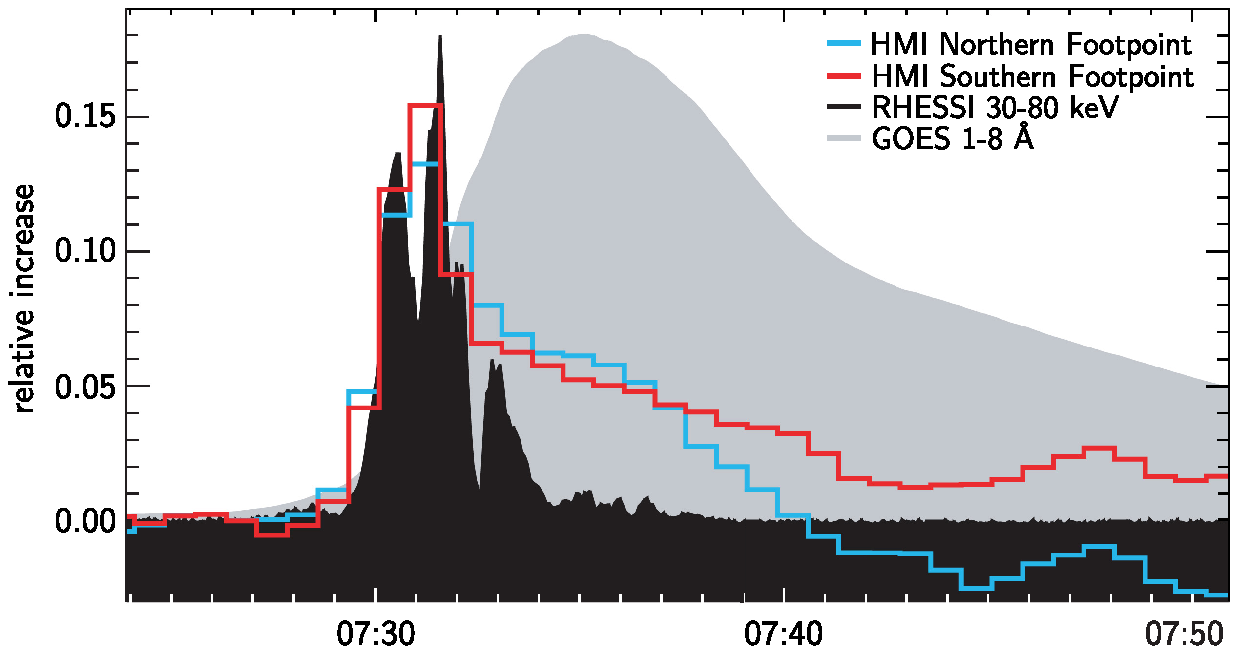}
\caption{Time series for SOL2011-02-24, showing the \textit{GOES} 1-8~\AA~soft X-rays in gray and the \textit{RHESSI} 30-80~keV flux in black.
The blue and red histograms show the mean white-light contrasts relative to the reference image (average between 07:25\,--\,07:28~UT), integrated over the footpoint areas shown in Figures~\ref{fig:footpoints}b and~\ref{fig:footpoints}c, respectively. 
The gradual mismatch of the two white-light footpoints results at least partly from the base-difference technique, and does not accurately represent flare emission. The lightcurves were scaled to fit on the plot.
}\label{fig:ts}
\end{figure}

The emission time series (Figure~\ref{fig:ts}) shows the timing behavior of hard X-ray and soft X-ray emissions typical of white-light flares \citep[e.g.][]{1975SoPh...40..141R,2010ApJ...715..651W}, and (again typically) we find a close match between the hard X-ray and white-light variations \citep[e.g.,][]{2011ApJ...739...96K}.
The time series show white-light differences relative to a pre-flare reference interval (07:25-07:28~UT).
The gradually increasing disagreement between the two footpoint light curves is the behavior expected for base-difference images in the presence of normal photospheric variability, which has time scales of minutes due to p-modes and granulation.

We determined the locations of the hard X-ray (HXR) sources by using the RHESSI visibilities-based forward fitting procedure, using data for the same 45-s time interval as observed by HMI.
We used a single 30-80~keV energy interval for simplicity and assumed a circular Gaussian model for each of the two bright sources.
The application of least-squares fitting then yielded the best-fit centroid coordinates of the two footpoints with statistical errors.
These positions have statistical errors $\approx$~0.16$''$.
Systematic errors in these positions come only from the RHESSI telescope metrology, which we believe to be substantially better than the statistical errors \citep{2003SPIE.4853...41Z}.
Since the uncertainty of the centroid position of the forward-fit model is smaller than its fitted size, there is the possibility
of systematic error if the true source shape is not well-approximated by a circular Gaussian.
Thus, in addition to the forward-fit image locations, we have separately characterized the source positions via the distribution of CLEAN component sources, which does not require the assumption of a specific model.
This check shows excellent consistency for the source centroids.

For the position of the white-light continuum sources, we have used the full-disk HMI image as a reference, fitting its limb via a standard inflection-point method.
Since the flare occurred near the limb, this greatly reduces the error in radial position due to any uncertainty in the HMI image properties, such as platescale.
We adopt the limb correction described by \cite{1998ApJ...500L.195B} to relate this measurement to the height of the photosphere (R$_\odot$); the mean of their two methods gives 498~km for the altitude separation, and we adopt 18~km to represent the uncertainty of this number.
At the time of the observation described here, SDO was near its antisolar point, and we have made the correction
to geocentric perspective.

\section{Relative source heights}\label{sec:relative}

The images in Figure~\ref{fig:footpoints} show good agreement in the radial coordinates of the source centroids in hard X-rays and white light.
We have made a small \textit{ad hoc} roll adjustment for the HMI observations, which has no effect on the source heights
because of the foreshortening.

Table~\ref{tab:big} summarizes the data from the three spacecraft and our estimates of uncertainties.
The STEREO entries characterize the heliographic (Stonyhurst) coordinates of the two footpoint sources, and serve to locate the projected position of the sub-flare photosphere on the HMI and RHESSI images.
The RHESSI positions include an estimate of the apparent source displacement due to scattering albedo,
based on the Monte Carlo simulations of \citet{2010A&A...513L...2K}.
In our observations, the primary and albedo sources would merge together, with a centroid slightly lower (closer to disk center) than the position of the primary source \citep[see Figure~4 in][for an example]{2011ApJ...739...96K}.
The anisotropy of the primary emission is likely to be small; in principle large anisotropies could lead to
larger albedo corrections \citep{2010A&A...513L...2K}.
At the extreme limb these simulations are not complete, since they do not incorporate photospheric rough structure \citep[e.g.,][]{1969SoPh....9..317S,1976ApJ...203..753L,1994IAUS..154..139C}.

This co-alignment analysis differs from that in Fig. 2 of \cite{2011A&A...533L...2B}, as the result of a different interpretation of the image metadata, and the overlay shown here is correct (M. Battaglia \& E. Kontar, personal communication 2011). 

\begin{table}
\caption{Source  Positions}
{\footnotesize
\bigskip
\begin{tabular}{| l | l | l | r | r | r | r |}
\hline
Quantity & Type & Units & Value & Err & Value & Err \\
\hline
\multicolumn{3}{| c |}{STEREO$^a$} & \multicolumn{2}{|c|}{X/Lon/EW}  & \multicolumn{2} {|c|} {Y/Lat/NS} \\
\hline
N source$^b$         & Meas  & px  &  1118.8 & 1.4 & 1206.5 & 1.0 \\
S source$^c$         & Meas   & px   &  1122.7  & 1.1 & 1198.5 & 1.0 \\
N Helio         & Calc   & deg  & 276.33 & 0.14  & 16.10 & 0.02\\
S Helio          & Calc  & deg          & 276.56 & 0.11 & 15.25 & 0.02\\
N Geo          & Calc   & arc s   & $-926.03$ & 0.15 & 279.53 & 0.05\\
S Geo          & Calc   & arc s  & $-929.43$ & 0.14  & 266.41& 0.04\\
\hline
\multicolumn{3}{| c |}{HMI}  & \multicolumn{2}{|c|}{X}  & \multicolumn{2} {|c|} {Y} \\
\hline
N Geo                              &   Meas  &     arc s     & $-926.76$ & 0.10 &  278.22 & 0.10 \\
S Geo                             &   Meas  &     arc s     & $-931.06$ & 0.10 & 261.47 & 0.10 \\
\hline
\multicolumn{3}{| c |}{RHESSI} & \multicolumn{2}{|c|}{X}  & \multicolumn{2} {|c|} {Y} \\
\hline
N Geo       &Meas      & arc s   & $-926.93$ & 0.16   & 278.06& 0.16 \\
S Geo        &  Meas    &  arc s    & $-931.19$  & 0.16  & 260.76 & 0.16 \\
\hline
\end{tabular}\label{tab:big}
}
\smallskip

$^a$ Sun center [1035.77, 1051.13] px (measured)

$^b$ Y pixel range 1205-1208

$^c$ Y pixel range 1196-1201
\end{table}

\section{Absolute source heights}\label{sec:absolute}

\begin{figure}[htbp]
\includegraphics[width=0.95\columnwidth, trim = 30 40 90 70]{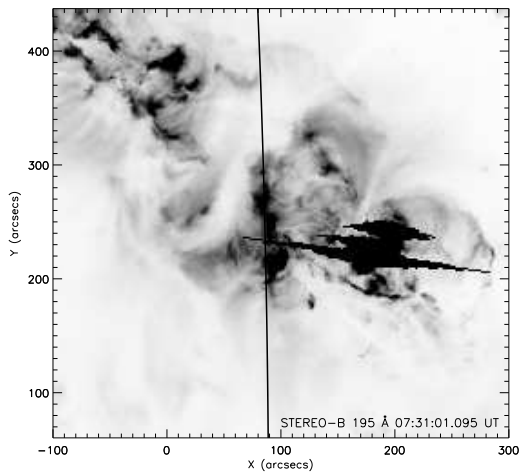}
\caption{The STEREO-B/EUVI image in the 195~\AA~band taken at 07:31:01.095~UT.
The flare occurs near disk center and results in image saturation as shown. 
The line shows the geocentric limb, demonstrating that the flare was not occulted.
}\label{fig:stereo_b}
\end{figure}

The proximity of the flare to the limb means that we can compare its central distance (the root-sum-square of its angular position coordinates) to the projected position of the photosphere, at the flare's heliographic location, to estimate the absolute source height.
We were fortunate in that this flare was observed near disk center by  \textit{STEREO-B}, then at a spacecraft heliolongitude of  $-$94.5$^\circ$.
The heliographic coordinates of the flare footpoints cannot be well determined from a geocentric perspective (\textit{RHESSI,  SDO}) because of the extreme foreshortening, but this is minimal for the \textit{STEREO-B} view.
One EUV (195~\AA) image (Figure~\ref{fig:stereo_b}) was taken during the integration time of the WL image shown in Figure~\ref{fig:footpoints}.
This 8-s EUVI image was saturated in about 16 of the columns of the CCD containing the flare itself.
The bright hard X-ray sources correspond to the southern two of the saturation regions, and a fainter HXR source, not detected in white light, lies to the north of these footpoint sources.

\begin{figure}[htdp]
\includegraphics[width=0.47\textwidth]{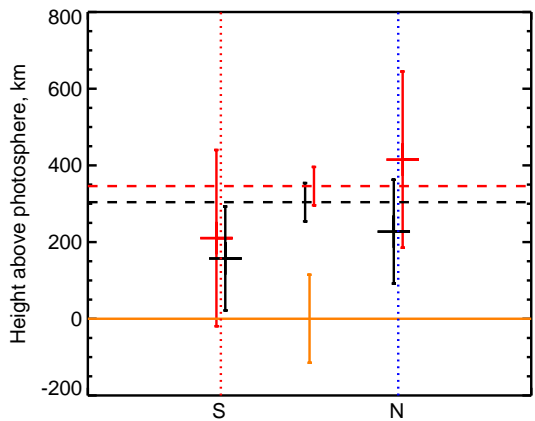}
\caption{Schematic view of the source heights determined for the centroids of the two footpoint sources of SOL2011-02-24.
Red points with errors show the hard X-ray source centroids ($>30$~keV), and black points the white-light sources.
The horizontal solid line with error bar is the zero point of the height measurement, set at the projected position of the corresponding STEREO sources.
The two dashed lines show the $\tau = 1$ points for Compton scattering opacity (red) and a scaled optical opacity (black).
The heights and their uncertainties for the (N, S) footpoints are $4.2 (2.1) \pm 2.4 \times 10^2$~km for HXR, and 
$2.3 (1.6) \pm 1.0 \times 10^2$~km for WL.
}
\label{fig:arrows}
\end{figure}

On each column of the CCD that is saturated, the excess charge should spread equally in both directions (M. Waltham, personal communication 2011).
Thus, to a first approximation, the mean row on each column (horizontal in the raw image) give the centroid of the image brightness on that column.
For the set of columns associated with each of the footpoints, we therefore estimate the heliographic coordinates and their uncertainties (Table~\ref{tab:big}) from the scatter of the data.
These coordinates agree, to within a few arcsec, with the positions of the WL and HXR sources (see Figure~\ref{fig:footpoints}).
The component of the uncertainty in the radial direction is small, as described below in the height measurement, because of the foreshortening.
The geometrical assumption here is that the heliographic coordinates of the EUV and hard X-ray sources coincide,
as they do in the impulsive phase of a flare \citep[e.g.,][]{2001SoPh..204...69F,2011ApJ...739...96K}.
We also implicitly assume that the EUV source is at zero height; we checked the uncertainty resulting from this systematic term and incorporate it in the height measurements. The result of these calculations give us a height for the Northern and southern HXR sources of $4.2 \times 10^2$ and $2.1 \times 10^2$~km respectively with an uncertainty of $2.4 \times 10^2$~km in both cases. The height of the intensity continuum sources was of $2.3 \times 10^2$ and $1.6 \times 10^2$~km for the Northern and southern sources, with uncertainties of $1.0 \times 10^2$~km for both measurements. These results are shown schematically in Figure~\ref{fig:arrows}. 

We note several other unknowns present in this procedure: the EUV and HXR sources may not actually have coincided in heliographic position; the images are not exactly cotemporal, since the EUV image has only an 8-s exposure time, so source variability could contribute to misalignment; finally, although impulsive-phase EUV and HXR sources can coincide precisely, some do not \citep[see references in][]{2011SSRv..159...19F}.
We mention these items for completeness but note that the extreme foreshortening minimizes their significance in this analysis.

Figure~\ref{fig:arrows} shows the computed height of unit optical depth to Compton scattering at about 350~km.
We estimated this from the 1D semi-empirical models of \cite{2009ApJ...707..482F}, taking the Compton cross-section
at 40~keV, a slightly higher energy than the 30~keV threshold for our HXR imaging.
In such a simple model atmosphere, we would not expect appreciable HXR emission to be detectable.
Note that this consideration would not affect the albedo source expected from a higher-altitude source, such as that implied by the thick-target model.
Figure~\ref{fig:arrows} also shows an estimate of the height of unit optical depth for the white-light continuum, derived here just
as a scaling from the quantities at the limb.

\section{Conclusions}\label{sec:concl}

In this study we have compared hard X-ray and white-light observations of a limb flare, SOL201-02-24T07:35:00.
The relative positions of the sources agree well, for each of the double-footpoint sources; since the local vertical maps almost exactly onto the solar radial coordinate, this means that the source heights match well.
Our uncertainties on this centroid matching are of order 0.2$''$.
This result strongly associates the white-light continuum with the collisional losses of the non-thermal electrons observed via bremsstrahlung hard X-rays in the impulsive phase of the flare.

We have also used the EUVI data in the 195~\AA~band from \textit{STEREO-B} to determine the heliographic coordinates of the flare footpoints, from a near-vertical vantage point.
To our knowledge, this enables the first direct determination of the absolute height of a white-light flare and its associated hard X-ray sources. Surprisingly, our estimates lie close to (if not below) the projected heights of optical depth unity for both 40~keV hard X-rays and for optical continuum at 6173~\AA.
They also lie well below the expected penetration depth of the $\sim$50-keV electrons needed to produce the hard X-rays (about 800~km for the Fontenla et al. quiet-Sun atmosphere, vs. 200-400~km as observed).
At present we have no explanation for this striking result and do not speculate about it, because it depends upon only a
single flare event.
We are sure that there are other comparable events in the existing data and hope to see a generalization of these results based on similar analyses.

\bigskip\noindent
{\bf Acknowledgements:} This work was supported by NASA under Contract NAS5-98033 for \textit{RHESSI} for authors Hudson, Hurford, Krucker, Lin, and Mart{\' i}nez Oliveros. R. Lin was also supported in part by the WCU grant (R31-10016) funded by the Korean Ministry of Education, Science, and Technology. Jesper Schou and Sebastien Couvidat are supported by NASA contract NAS5-02139 to Stanford University. The HMI data used are courtesy of NASA/SDO and the HMI science team.
We thank Martin Fivian for helpful discussions of the RHESSI aspect system.
Alex Zehnder's precise metrology of RHESSI has made this analysis possible in the first place.
We further thank M. Waltham for comments on the saturation properties of the SECCHI CCD detectors.

\nocite{2011A&A...533L...2B}
\nocite{1971SoPh...18..489B}
\nocite{1859MNRAs..20...13C}
\nocite{1969sfsr.conf..356E}
\nocite{2005JGRA..11011103E}
\nocite{2001SoPh..204...69F}
\nocite{1859MNRAs..20...16H}
\nocite{1972SoPh...24..414H}
\nocite{1992PASJ...44L..77H}
\nocite{1976SoPh...50..153L}
\nocite{2003ApJ...595L..69L}
\nocite{1976ApJ...203..753L}
\nocite{1989SoPh..124..303M}
\nocite{1970SoPh...15..176N}
\nocite{1989SoPh..121..261N}
\nocite{1993SoPh..143..201N}
\nocite{1975SoPh...40..141R}
\nocite{1970SoPh...13..471S}
\nocite{1981ApJS...45..635V}
\nocite{2009RAA.....9..127W}
\nocite{2010ApJ...715..651W}
\nocite{1998ApJ...500L.195B}

\bibliographystyle{apj}

%\date{\today}

\end{document}